\documentstyle[fleqn,psfig]{revp} 
\newfont{\sff}{cmssi12} 
\newfont{\bigsf}{cmss12 scaled 2000} 
\newfont{\midsf}{cmss12 scaled 1000} 
\newfont{\smlsf}{cmss12 scaled 600}  
\newfont{\bigsff}{cmssi12 scaled 2000} 
\newfont{\sfi}{cmssi10} 

\newcommand{\pmonth}{M\"arz, 1999}
\newcommand{\spec}{nuclear physics: NUCOLEX99 20.03.-24.03.1999}
\newcommand{\no}{} 

\newcommand{\be}{\begin{eqnarray}}
\newcommand{\ee}{\end{eqnarray}}

\begin{document}
\parindent 0pt
\parskip 12pt
\setcounter{page}{1}

\title{ISGDR- bulk or surface mode ?}

\author{Klaus Morawetz,$^{*1}$ Uwe Fuhrmann,$^{*1}$ \\  
\sff
*1 Fachbereich Physik, University Rostock, D-18055 Rostock,
Germany\\}

\abst{
A polarization and response function for finite systems and temperatures is 
derived from linearizing the Vlasov equation. Besides the Lindhard response 
function in local density approximation we obtain an additional contribution 
due to the surface. This formula is applied to isoscalar giant resonances 
which we consider as next harmonics to the basic dipole mode. The spurious 
center of mass motion is subtracted within the polarization. 
The results based on simple Fermi liquid considerations are in 
reasonable agreement with experimental data. 
}

\maketitle
\thispagestyle{headings}

The existence of isoscalar giant dipole resonance (ISGDR) in nuclear
matter is considered as a spurious mode in most text books since
one associates with it a center of mass motion. The more surprising
was the experimental justification of a giant resonance carrying
the quantum numbers of a isoscalar and dipole mode$^{1)2)}$. 
Consequently
one has to consider higher harmonics as explanation of such a
mode$^{1)3)4)}$. 
Usually this mode is associated with a squeezing mode analogous 
to a sound wave$^{1)2)5)}$.

In this letter we want to discuss the influence of surface effects on
the ISGDR compression mode. We will show that even in the frame 
of the Fermi liquid model such mode can be understood. 
Moreover we claim that the surface
effects are not negligible for reproducing the strength function.
Consequently we give first a short derivation of response
function including surface effects. This will result into a new
formula on the level of temperature dependent extended Thomas
Fermi approximation$^{6)}$.

The starting point is the semiclasssical Vlasov equation
\be
\partial_t f+{{\bf p}\over m}{\bf \nabla}_{R} {\bf \nabla}_p
f-{\bf \nabla}_R(U^{\rm ext}+U^{\rm ind}){\bf \nabla}_pf=0
\label{vlasov}
\ee
where $U^{\rm ext}$ is the external perturbation and $U^{\rm ind}$ 
the selfconsistent meanfield. Provided we know the response
to the external potential without selfconsistent meanfield, which
is described by the polarization function $\Pi$
\be
\delta n({\bf x},t)=\int d{\bf x}' \Pi({\bf x,x'},t) U^{\rm
ext}({\bf x'},t).
\label{po}
\ee
The response including meanfield $\chi$, is given by
\be
&&\chi({\bf x,x'},t)=\Pi({\bf x,x'},t)\nonumber\\
&&+\int d{\bf x}_1d{\bf x}_2
\Pi({\bf x},{\bf x}_1,t) {\delta
U^{\rm ind}({\bf x}_1,t)\over \delta n({\bf x}_2,t)} \chi({\bf
x}_2,{\bf x'},t).
\label{chi}
\ee

Therefore we concentrate first on the calculation of the
polarization function $\Pi$ and linearize the
Vlasov equation (\ref{vlasov}) according to 
\be
f({\bf
p,R},t)=f_0({\bf p,R})+\delta f({\bf p,R},t)
\ee 
such that the
induced density variation
$\delta n({\bf R},t)=$\\$\int \frac{d{\bf p}}
{(2 \pi \hbar)^3}\delta f$ reads
\be
&&\delta n({\bf x},\omega)=\int{d{\bf q}\over (2\pi\hbar)^3} {\rm e}^{i{\bf
qx}}\nonumber\\
&&\times\int{d{\bf p}d{\bf x'}\over (2\pi\hbar)^6}{\rm e}^{-i{\bf qx'}}
{{\bf \nabla}_p
f_0({\bf p,x'}) {\bf \nabla}_{x'} U^{\rm ext}({\bf x'},t)
\over i(\omega-{{\bf pq}\over m})}.\nonumber\\
&&
\label{1}
\ee
Here we have employed the Fouriertransform of space and time coordinates
of (\ref{vlasov}) to solve for $\delta f$ and inverse
transform the momentum into the form (\ref{1}).
Comparing (\ref{1}) with the definition
of the polarization function (\ref{po}) we extract with one
partial integration
\be
\lefteqn{\Pi({\bf x,x'},\omega)=-{\bf \nabla}_{x'}\int{d{\bf p}d{\bf q}\over 
(2\pi\hbar)^6}
{\rm e}^{i{\bf q} ({\bf x-x'})}{{\bf \nabla}_p f_0({\bf p,x'})
\over i (\omega-{{\bf pq}\over m})}. } 
\label{p}
\ee
With (\ref{p}) and (\ref{chi}) we have given the polarization and
response functions for a finite system.

In the following we are interested in the gradient expansion
since we believe that the first order gradient terms will bear
the information about surface effects. Therefore we change to
center of mass and difference coordinates ${\bf R}=({\bf
x}_1+{\bf x}_2)/2$,
${\bf r}={\bf x}_1-{\bf x}_2$ and retaining only first order
gradients we get from (\ref{p}) after Fourier transform of ${\bf r}$
into ${\bf q}$
\be
\lefteqn{\Pi({\bf R,q})=-\int{d{\bf p}\over 
(2\pi\hbar)^3}{{\bf q \nabla}_p f_0({\bf p,R})\over \omega-{{\bf p q}\over m}} }
\nonumber\\
\lefteqn{+{i\over 2} {\bf \nabla}_R
\int{d{\bf p}\over (2\pi\hbar)^3} \biggl ( {{\bf \nabla}_p f_0({\bf p,R})\over
\omega-{{\bf pq}\over m}}-{{\bf p}\over m} {{\bf q.\nabla}_p
f_0({\bf p,R})\over (\omega-{{\bf pq}\over m})^2} \biggr )} \nonumber\\
\lefteqn{=-\int{d{\bf p}\over (2\pi\hbar)^3} [{\bf q}-i{\bf \nabla}_R 
(1+{\omega\over 2}\partial_{\omega})]{{\bf \nabla}_p f_0({p^2\over 2 m},
{\bf R})\over\omega-{{\bf pq}\over m}} }
\label{grad}
\ee
where in the last equality we have assumed radial momentum
dependence of the distribution function $f_0$. We recognize that
besides the usual Lindhard polarization function as the first
part of (\ref{grad}) we obtain a second part which is expressed
by a gradient in space. The first part corresponds to the Thomas
Fermi result where we have to use the spatial dependence in the
distribution functions and the second part represents the
extended Thomas Fermi approximation. So far we did not assume any
special form of the distribution function. Therefore the
expression (\ref{grad}) is as well valid for any high temperature
polarization of finite systems.

What remains is to show that the response function (\ref{chi})
does not contain additional gradients. This is easily confirmed
by two equivalent formulations of (\ref{chi}),  $\Pi^{-1}\chi=1+V\chi$ and
$\chi\Pi^{-1}=1+\chi V$, which by adding yield the anticommutator
\be
[\Pi^{-1},\chi]_+=2+[V,\chi]_+.
\ee
This anticommutator does not contain any gradients up to second order. 
Therefore we have [$V={\delta U^{\rm ind}/\delta n}$]
\be
 \chi({\bf R,q},\omega)={\Pi({\bf R,q},\omega)\over 1-V({\bf R,q},\omega)
\Pi({\bf R,q},\omega)} +{\cal O}(\partial_R^2). \label{c}
\ee
Equation (\ref{c}) and (\ref{grad}) give the response and polarization 
functions of finite systems in first order gradient approximation.

Now we are ready to derive approximate formulae for spherical
nuclei. In this case we can assume ${\bf q}||{\bf R}$ and we have
\be
\lefteqn{\Pi(R,{\bf q},\omega)=\Pi^0(R,{\bf q},\omega)-{i\over q} 
\partial_R \left[1+{\omega\over
2}\partial_{\omega}\right] \Pi^0(R,{\bf q},\omega)}\nonumber\\ \label{2}
\ee
where $\Pi_0$ is the usual Lindhard polarization with spatial
dependent distributions (chemical potentials, density).
We use now further approximations. In the case
of giant resonances we are in the regime where ${\rm Im}\Pi^0 \sim
\omega$ such that 
\be
\left[1+{\omega\over
2}\partial_{\omega}\right] \Pi^0=i \frac 3 2 {\rm Im} \Pi^0
\label{10}
\ee
and for small $q$
the real part vanishes for ${\rm Re}\Pi^0\sim 1-c^2q^2/\omega^2$. 
Within the local density approximation we know that the spatial dependence is
due to the density $n(R)=n_0\Theta(R_0-R)$. Since we have for zero temperature 
${\rm Im} \Pi^0\propto p_f(n)$ we evaluate
\be
\partial_R {\rm Im}\Pi^0&=&-n_0 \delta(R_0-R)\partial_n {\rm Im}\Pi^0
\nonumber\\
&=&-\frac 1 3 \delta(R_0-R) {\rm Im}\Pi^0
\label{11}
\ee
where we assumed the density dependence carried only by the
Fermi momentum.
Now it is straight forward to spatially average (\ref{2})
with the help of (\ref{11})
\be
\Pi({\bf q},\omega)&=&{3\over R_0^3} \int\limits_0^{R_0} d R R^2
\Pi(R,{\bf q},\omega)\nonumber\\
&\approx&\Pi^0({\bf q},\omega)+ i {3\over q R_0} \left[1+{\omega\over
2}\partial_{\omega}\right] \frac 1 3 \Pi^0({\bf q},\omega)
\nonumber\\
&=&\Pi^0+\Pi^{\rm surf}.
\label{3}
\ee
Consequently the surface contribution to the
polarization function reads finally with (\ref{10})
\be
\Pi^{\rm surf}({\bf q},\omega)=-{3\over 2 q R_0}{\rm Im}\Pi^0({\bf q},\omega)
\label{4}
\ee
which is real. With (\ref{3}), (\ref{4}) and (\ref{c}) we obtain
finally for the structure function 
\be
S(q,\omega)=\frac 1 \pi {{\rm Im}\Pi^0\over\left[1-V ({\rm Re} \Pi^0+\Pi^{\rm
surf})\right]^2+(V{\rm Im}\Pi^0)^2}.
\label{s}
\ee
For small $q$ expansion we see that the pole of the structure function
becomes renormalized similar as known from the Mie mode or
surface plasmon mode$^{7)8)}$
\be
\omega^2={\omega_0^2\over 1-V \Pi^{\rm
surf}}.
\ee

\begin{figure}[h]
\psfig{file=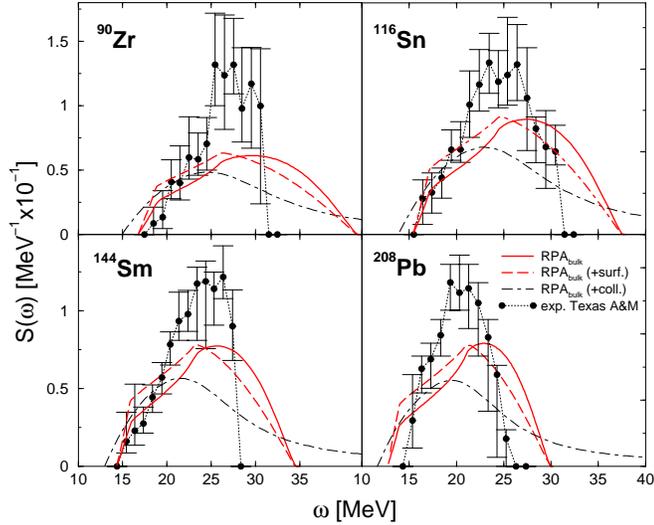,width=9cm,angle=-90}
\caption{The experimental structure function (T=0) 
versus theoretical values. 
The bulk RPA result (solid lines) is compared with the 
extended Thomas Fermi approximation (surface corrections, dashed lines)     
and the inclusion of collisions (dot-dashed lines). 
The latter one should be of less 
importance due to symmetry of isoscalar mode. The data suggest this case and 
support surface contributions. Circles: Normalized data from Ref.$^{14)}$}
\end{figure}

After establishing the structure function including surface
contribution we specify the model for actual calculations. We
choose as mean field parameterization a Skyrme force following 
Vautherin and Brink$^{9)}$ which leads to the isoscalar potential
\be
V=\frac{3t_0}{4}+\frac{3t_3}{8}n_0
\ee
with $t_0=-983.4$ MeV fm$^3$, $t_3=13106$ MeV fm$^6$, $x_0=0.48$ at
nuclear saturation density $n_0=0.16$ fm$^{-3}$ and
the incompressibility of $K=318$ MeV. 
Further we employ the Steinwedel-Jensen model$^{10)}$ 
where the basic mode inside a sphere of radius $R_0$ is given by a wave vector
\be
q_{sp}={\pi\over 2 R_0}.
\ee
This would correspond to the first order isovector mode$^{11)}$. 
Since this mode is spurious$^{12)}$ we have to consider the next higher
harmonics$^{13)}$ which is
\be
q_{isgdr}={\pi\over R_0}.
\ee
Since the polarization function with this second order mode contains
still contributions from the spurious mode we have to subtract
this part$^{3)4)}$
\be
\Pi^0_{\rm ISGDR}(\omega)=\Pi^0(q_{isgdr},\omega)-\Pi^0(q_{sp},\omega).
\label{5}
\ee
In Fig.1 we have plotted the 
experimental structure function together 
with different theoretical estimates according to (\ref{5}) and (\ref{s}). The 
inclusion 
of surface corrections (dashed lines) shifts the structure function 
towards the experimental 
values. The inclusion of collisions (dot-dashed lines), which should be 
of minor importance for 
isoscalar dipole mode due to cancellation of backscattering, leads 
to worse results.
The results support also that the mode is of isoscalar dipole type.

\section*{References}

\re
1) M.\ N.\ Harakeh and A.\ E.\ L.\ Dieperink: Phys.\ Rev.\ C\ {\bf 23}, 
2329  (1981).
\re
2) B.\ F.\ Davis et al.: Phys.\ Rev.\ Lett.\ {\bf 79},  609  (1997).
\re
3) T.\ J.\ Deal: Nucl.\ Phys.\ {\bf A217}, 210  (1973).
\re
4) I.\ Hamamoto,\ H.\ Sagawa,\ and X.\ Z.\ Zhang: Phys.\ Rev.\ C\ {\bf 57}, 
R1064  (1998).
\re
5) N.\ V.\ Gai and H.\ Sagawa: Nucl.\ Phys.\ {\bf A371}, 1 (1981).
\re
6) P.\ Ring and P.\ Schuck: {\em The Nuclear Many-Body Problem} (Springer-Verlag, New
  York, 1980).
\re
7) G.\ F.\ Bertsch and R.\ A.\ Broglia: {\em Oscillations in Finite Quantum Systems} 
(Cambridge Mongraphs, New York, 1994).
\re
8) U.\ Kreibig and M.\ Vollmer: {\em Optical Properties of Metal Cluster} 
(Springer-Verlag,
Berlin, 1995).
\re
9) D.\ Vautherin and D.\ M.\ Brink: Phys.\ Rev.\ C\ {\bf 5},  626  (1972).
\re
10) H.\ Steinwedel and J.\ Jensen: Z.\ f.\ Naturforschung\ {\bf 5},  
413  (1950).
\re
11) F.\ Braghin and D.\ Vautherin: Phys.\ Lett.\ B\ {\bf 333},  289  (1994).
\re
12) K.\ Morawetz, R.\ Walke, and U.\ Fuhrmann: Phys.\ Rev.\ C\ {\bf 58}, 
1473 (1998).
\re
13) K.\ Morawetz, U.\ Fuhrmann, and R.\ Walke: Nucl.\ Phys.\ {\bf A649}, 
348 (1999).
\re
14) U.\ Garg: private communication.


\end{document}